\documentclass[letterpaper, 10 pt, conference]{ieeeconf}  

\IEEEoverridecommandlockouts                              
\usepackage{graphicx} 
\usepackage{amsmath} 
\usepackage{amssymb}  
\usepackage{textcomp}
\usepackage{makecell}
\usepackage{hhline}
\usepackage{algorithm}
\usepackage{algorithmic}
\usepackage{bbm}
\usepackage{caption} 
\usepackage{subcaption} 
\usepackage{color} 
\usepackage{hyperref}
\usepackage{multirow}
\usepackage{dsfont}
\usepackage[tableposition = top]{caption}

\title{\LARGE \bf
Nested Reinforcement Learning Based Control  for Protective Relays  in Power Distribution Systems
}

\author{Dongqi Wu, Xiangtian Zheng, Dileep  Kalathil, Le Xie
\thanks{The authors are with the Department of Electrical and Computer Engineering, Texas A\&M University, Texas, United States. Email: {\tt\small \{dqwu, zxt0515, dileep.kalathil, le.xie\}@tamu.edu}}}

\begin{document}

\maketitle
\thispagestyle{empty}
\pagestyle{empty}

\begin{abstract}
This paper envisions a new control architecture for the protective relay setting in future power distribution systems. With deepening penetration of distributed energy resources at the end users level, it has been recognized as a key engineering challenge to redesign the protective relays in the future distribution system. Conceptually, these protective relays are the discrete ON/OFF control devices at the end of each branch and node in a power network. The key technical difficulty lies in how to set up the relay control logic so that the protection could successfully differentiate heavy load and faulty operating conditions. This paper proposes a new nested reinforcement learning approach to take advantage of the structural properties of distribution networks and develop a new set of training methods for tuning the protective relays.

%
%
%

\end{abstract}


\section{INTRODUCTION} 

This paper is motivated by the increasing need to re-design the control architecture of protective relays in the power distribution systems. The goal of protective relays is to detect abnormal conditions, such as short circuit and equipment failures, and isolate the corresponding elements to prevent possible cascading destruction. The key design criteria for protective relays in the power distribution system is to properly isolate faults under abnormal conditions while not tripping under normal operating conditions. Since the protective relays are installed at all the nodes and branches, tripping of a protective relay would have consequences beyond the immediate neighboring device in the system. Therefore, the art and science of designing a protective relay system lies in how to trade-off different protective relay tripping during faulty situations. With increasing level of uncertainties in line flow patterns due to distributed energy resources, the design of a intelligent relay system has become the key engineering challenge to fully realize the potential of a truly low-carbon energy system in the future. This paper directly addresses this challenge of how to re-design the protective relay systems in the distribution grid. 

This paper focuses on the re-design of the control logic for overcurrent relays. Overcurrent relays are the most widely used protective relays in the power grid. Overcurrent relays use the current magnitude as the indicator of faults. When a short-circuit fault occurs, the fault current is typically much larger than the nominal current under the normal conditions. The operating principle of this kind of relay is to trip the line if the measured current exceeds a pre-fixed threshold. This threshold is usually determined based on a number of heuristics that account for the topology of the network and feeder capacity.   

In the case of possible operation failure of any relay, some coordination between adjacent relays is necessary to avoid catastrophic outcomes. This is typically achieved with a primary relay - backup relay coordination. If a faults occurs in the assigned region of a given relay, it should act as the primary relay and trip. If (and only if) the primary relay fails to trip, the adjacent upstream (towards the feeder) relay should trip. Since there is no explicit communication between the relays, this coordination is achieved implicitly using an  `inverse time curve' \cite{PRBook}. The basic idea is to design relays in such a way that higher fault current  results in shorter tripping time. Since the primary relay is closer to the fault, the measured current will be higher and it will trip faster. If the primary relay fails, the backup relay will work but only after some time delay indicated by the inverse time curve.

Successful operation of conventional overcurrent relays rely on two crucial assumptions: (i) nominal operation currents are always less than the fault current, (ii) the current measurements are always higher for the relays that are closer to the fault. With the increasing penetration of distributed energy resources, both assumptions are likely to be rendered invalid. This is due to the fact that distributed energy resources such as solar panels and batteries may create reverse power flows from the edge of the grid to the substation. 

An efficient control algorithm for relay protection should be able to:  (i) reduce the operation failures as low as possible,  (ii) identify the fault as soon as possible, and (iii) adapt robustly against the changes in the operating conditions, like shift in the load profile. A unified approach that can exploit the availability of huge amounts of real-time sensor data from the power distribution systems, recent advances in machine learning, along with domain knowledge of the power systems operations is necessary to achieve these objective, especially in the context of next generation power systems.  

{\bf Related work:}  Most studies of improving the performance of the over-current relays focus on the aspects of coordination \cite{mlr1} \cite{mlr2}, fault detection \cite{mlr3} and fault section estimation \cite{mlr4}. Among various possible methods, machine learning is popular for advanced over-current relays. Neural networks \cite{mlr5} \cite{mlr6} are applied to determine the coefficients of the inverse-time over-current curve. Other research work based on support vector machine \cite{mlr7} \cite{mlr8} directly determine the operation of relays. However, most of these learning techniques do not explicitly explore the dynamic nature of the protective relay setting. As the power network grows in its complexity and flow patterns, it is often difficult to differentiate normal setting from a faulty one simply from a snapshot of measurements.

Reinforcement learning (RL) is a class of machine learning that focuses on \textit{learning to control}  unknown dynamical systems. Unlike the other two classes of machine learning,  supervised learning and unsupervised learning, which typically focus on static systems, RL  methodology explicitly includes the tools to characterize the dynamical nature of the system that it tries to learn.  In RL, a learning agent  sequentially  interacts with a (dynamical) system by observing the states of the system and  by selecting   appropriate control actions.  The system state evolves stochastically depending on the control action of the agent and according to the (unknown) model of the system. After each interaction, the agent  receives a reward.  The goal of the agent is to learn a control policy, which specifies the optimal  action to take given the state of the system, in order to to maximize the cumulative reward.

Last few years have seen significant progresses in deep neural netwoks based RL approaches for controlling unknown dynamical systems, with applications in many areas like playing games \cite{silver2016mastering}, locomotion \cite{lillicrap2015continuous} and robotic hand manipulation \cite{levine2016end}.  This  has also led to addressing many power systems problems using the tools from RL, as detailed in the survey \cite{Glavic17}.  RL is indeed the most appropriate machine learning approach for a large class of power systems problems because of the inherent stochastic and dynamical nature of the power systems.  RL based applications in power systems include electricity markets \cite{ems1} \cite{ems2}, voltage control \cite{vc1}, automatic generation control \cite{agc1}, demand control \cite{dc1}, and  angle stability control  \cite{tai1} \cite{oai1}. However, little effort has been made for using RL for relay protection control. The closest work  \cite{RL1} discusses about using a centralized Q-learning algorithm  to determine the protection strategy for a relay network with full communication between them. The prerequisite of global communication leaves this method impractical.     


{\bf Our contributions:} We propose a novel \textit{nested reinforcement learning} algorithm for optimal relay protection control for a network of relays  in a power distribution network. We don't assume any explicit communication between the relays. We formulate the relay protection control as a multi-agent RL problem where each relay acts as an agent, observes only its local measurements and takes control actions based on this observation. Multi-agent RL problems are known to be intractable in general and convergence results  are sparse. We overcome this difficulty by cleverly exploiting the underlying radial structure of power distribution systems. We argue that this structure imposes only a one directional influence pattern among the agents, starting from the end of the line to the feeder. Using this structure, we develop a nested training procedure for the network of relays. Unlike generic multi-agent RL algorithms which often exhibit osculations and even non-convergence in training, our nested RL algorithm converges fast in simulations. The converged policy far outperforms the conventional threshold based relay protection strategy in terms of failure rates, robustness to change in the operation conditions, and speed in responses. 

This paper is organized as follows. Section II formulates the relay operation problem. Section III gives a brief review on RL. Section IV provides our new algorithm.  Section V presents  simulation studies  that show the efficiency of the proposed method. Concluding remarks are presented in Section VI.

\section{Problem Formulation}

\begin{figure}[t]
\centering
\includegraphics[scale=0.3]{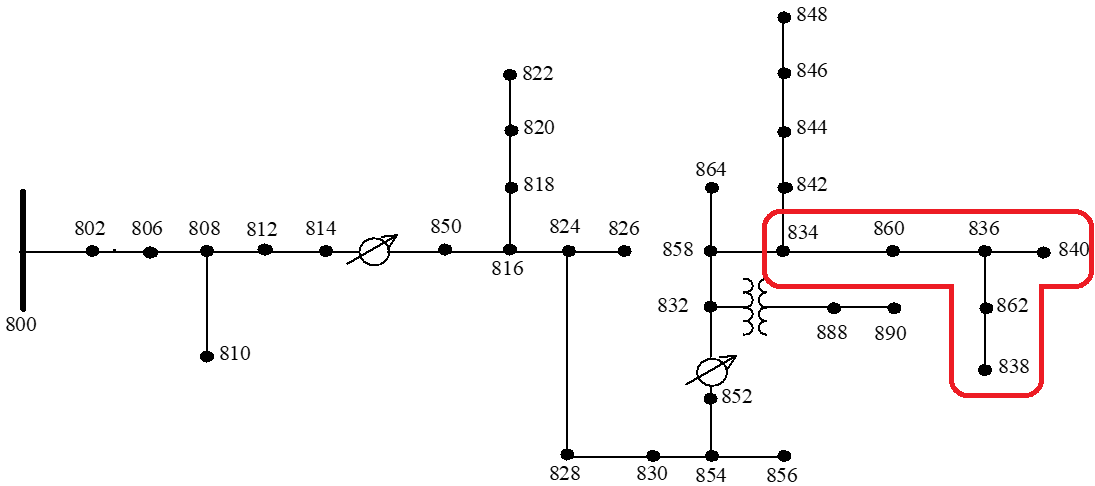}
\caption{IEEE34 node test feeder}
\label{fig:IEEE34}
\end{figure}
\begin{figure}[t]
\centering
\includegraphics[scale=0.35]{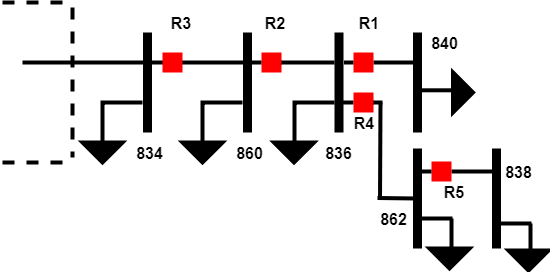}
\caption{Protective relays in a radial network}
\label{fig:radial1}
\end{figure}

In order to precisely characterize  the operation of protective relays, we first explain what ideal relays are supposed to do using a concrete setting given in Fig. \ref{fig:radial1}. This is a small section of the larger standard IEEE 34 node test feeder \cite{IEEE34} shown in Fig. \ref{fig:IEEE34}.  There are five relays  protecting five segments of the distribution line.  

Desirable operation of the relays is as follows. Each relay is located to the right of a bus (node). Each relay needs to protect its own region, which is between its own bus and the first down-stream bus. Relays  are also required  to provide backup for its first downstream neighbor: when its neighbor fails to operate, it needs to trip the line and clears the fault. For example, in Fig. \ref{fig:radial1}, if a fault occurs between bus 862 and 838, relay 5 is the main relay protecting this segment and it should trip the line immediately. If relay 5 fails to work, relay 4, which provides backup for relay 5, needs to trip the line instead. The time delay between detecting and clearing a fault should be as short as possible for primary relays, while backup relays should react slower to ensure that they are only triggered when the corresponding main relay is not working.

According to the above description,   the desired operation of a network of protective relays  can be formalized as follows.  Suppose there are $n$ relays.  Denote the control action of $i$th relay as $a_i$ and the index of its downstream neighbor as $n_i$. For $a_i$, 1 means to trip while 0 means to hold. Let $X_{p_i}$ and $X_{b_i}$ respectively be the primary and backup protection region of relay $i$. Suppose $x_f$ is the location of the fault, then the ideal control action $a_{i}$ of relay $i$ is
\begin{align}
a_i=\mathds{1}((x_f {\in} X_{P_i}) \cup((x_f {\in} X_{b_i})\cap (a_{n_i}=0)))
\end{align}
where $\mathds{1}(\cdot)$ is an indicator function.

However, in practice each relay knows only the local measurements like voltage and current. In particular, relay $i$ is not aware of downstream neighbors' actions $a_{n_{i}}$ and the exact location of the fault $x_{f}$. So, each relay $i$ needs a local control policy $\pi_{i}$ that maps the local observation $s_{i}$ to control action $a_{i}$, i.e., $a_{i} = \pi_{i}(s_{i})$. These local control polices are to be designed in such a way to enable an implicit coordination between the relays in the network to achieve the a global protection strategy.   In following, we propose a multi-agent reinforcement learning approach for addressing this problem.

\section{Markov Decision Processes and Reinforcement Learning}
Before formulating the relay protection problem using the RL approach, we first give a brief review of  some basic terminologies in RL. 

\emph{Markov decision processes} (MDP) is a canonical formalism for stochastic  control problems. The goal is to  solve sequential decision making (control) problems in stochastic  environments where the control actions can influence the evolution of the state of the system. An MDP is modeled as  tuple $(\mathcal{S}, \mathcal{A}, R, P, \gamma)$ where $\mathcal{S}$ is the state space, $\mathcal{A}$ is the action space. $P = (P(\cdot|s, a), (s, a) \in \mathcal{S} \times \mathcal{A})$ are the state tranistion probabilities. $P(s'|s, a)$ specifies the probability of transition to $s'$ upon taking action $a$ in state $s$. $R : \mathcal{S} \times \mathcal{A} \rightarrow \mathbb{R}$ is the reward function, and $\gamma \in [0, 1)$ is the discount factor. 

A policy $\pi : \mathcal{S} \rightarrow \mathcal{A}$ specifies the control action to take in each possible state. The performance of a policy is measured using the metric  \textit{value of a policy}, $V_{\pi}$, defined as 
\begin{align}
&V_{\pi}(s) = \mathbb{E}[\sum^{\infty}_{t=0} \gamma^{t} R_{t}   | s_{0} =s ],
\end{align}
where $R_{t} = R(s_{t}, a_{t}),  a_{t} = \pi(s_{t}), s_{t+1} \sim P(\cdot | s_{t}, a_{t})$. 
The optimal value function $V^{*}$ is defined as $V^{*}(s) = \max_{\pi} ~V_{\pi}(s)$. 
Given $V^{*}$, the optimal policy $\pi^{*}$ can be calculated using the Bellman equation as
\begin{align}
\pi^{*}(s) = \arg \max_{a \in \mathcal{A}} ~(R(s, a) + \gamma \sum_{s' \in \mathcal{S}} P(s'|s, a) V^{*}(s') ).
\end{align} 

Similar to the value function,  Q-value function of a policy $\pi$, $Q_{\pi}$, is defined as
\begin{align}
Q_{\pi}(s, a) = \mathbb{E} [\sum^{\infty}_{t=0} \gamma^{t} R_{t}  | s_{0} =s, a_{0} = a ]
\end{align} 
Optimal Q-value function $Q^{*}$ is also defined similarly, $Q^{*}(s, a) = \max_{\pi} Q_{\pi}(s, a)$. Optimal Q-value function  will help us to compute the optimal policy directly without using the Bellman equation, as $\pi^{*}(s) = \arg \max_{a \in \mathcal{A}} ~ Q^{*}(s, a)$ 

Given an MDP formulation, the optimal value/Q-value function ($V^{*}/Q^{*}$) 
or the optimal control policy ($\pi^{*}$) can be computed using dynamic programming  methods like value iteration or policy iteration \cite{sutton2018reinforcement}. However, these dynamic programming method requires the knowledge of the full model of the system, namely, the transition probability $P$ and reward function $R$. In most real world applications, the stochastic system model is either  unknown or extremely difficult to model.  In the protective relay problem, the transition probability  represents all the possible stochastic variations in voltage and current  in the network, due to a large number of scenarios like weather (and the resulting shift in demand/supply) and renewable energy generation. In such scenarios, the optimal policy has to be \textit{learned} from sequential state/reward observations.

\textit{Reinforcement learning}  is a method for computing the optimal policy for an MDP when the model is unknown. RL achieves this without explicitly constructing an empirical model. Instead, it directly learns the optimal Q-value function or optimal policy from the sequential observation of states and rewards.   

\textit{Q-learning}  is one of the most popular RL algorithms which learn the optimal  $Q^{*}$ from the sequence of observations $\left(s_t, a_t, R_{t}, s_{t+1} \right)$. Q-learning algorithm is implemented as follows. At each time step $t$, the RL agent updates the Q-function  $Q_{t}$ as  
\begin{align}
Q_{t+1}(s_{t},a_{t}) &= (1-\alpha_t)Q_t(s_{t},a_{t})  \nonumber \\
&\hspace{1cm} + \alpha_{t} ( R_{t} + \gamma \max_b Q_{t}(s_{t+1},b))
\end{align}
where $\alpha_{t}$ is the step size (learning rate). It is known that if each-state action pairs is sampled infinitely often and under some suitable conditions on the step size, $Q_{t}$ will converge to the optimal Q-function $Q^{*}$ \cite{sutton2018reinforcement}.

Using a standard tabular Q-learning algorithm as described above is  infeasible  in problems with continuous  state/action space. To address this problem, Q-function is typically approximated using a deep neural network, i.e., $Q(s,a) \approx Q_{w}(s,a)$ where $w$ is the parameter of the neural network. Deep neural networks  can approximate arbitrary functions without explicitly designing the features and this has enabled tremendous success in both supervised learning (image recognition, speech processing) and reinforcement learning (AlphaGo games) tasks.   

In Q-learning with neural network based approximation, the parameters of the neural network can be updated using stochastic gradient descent with step size $\alpha$ as  
\begin{align}
\label{eqn:sgd}
w &= w + \alpha \nabla Q_w(s_t,a_t) \nonumber  \\
&\hspace{1cm} (R_t + \gamma \max_b Q_w(s_{t+1},b) - Q_w(s_t,a_t))
\end{align}

Unlike supervised learning algorithm, the data samples $(s_t, a_t, R_t, s_{t+1})$ obtained by an RL algorithm is correlated  in time due to the underlying system dynamics. This often leads to a very slow convergence or non-convergence of the gradient descent algorithms like \eqref{eqn:sgd}. The idea of \textit{experience replay} is to break this temporal correlation by randomly sampling some data points from a buffer of previously observed (experienced) data points to perform the gradient step in \eqref{eqn:sgd}. New observations are then added to the replay buffer and the process is repeated.

In  the gradient descent equation \eqref{eqn:sgd}, the target  $R_t + \gamma \max_b Q_w(s_{t+1},b)$ depends on the neural network parameter $w$, unlike the targets used for supervised learning which are fixed before learning begins. This often leads to poor convergence in RL algorithms. The idea of \textit{target network} is used to address this issue.  A separate (target) neural network is used for maintaining the target value for gradient descent.  The target network is kept fixed for multiple steps but updated periodically.   


The combination of neural networks, experience replay and target network forms the core of the DQN algorithm \cite{DQN}.  In the following, we will use DQN  as one of the basic block for our nested RL algorithm.

\section{Nested Reinforcement Learning  for Control of Protective Relays}

We model the protective relays as collection of RL agents. Each relay can only observe its local measurements of voltage and current. Each relay also knows the status of the local current breaker circuits, i.e., if it is open or closed. Since relays don't observe the measurements at other relays, an implicit coordination mechanism is also needed in each relay. This is achieved by including a local counter that ensures the necessary time delay in its operation as backup relay. These variables constitute the state $s_{i}(t)$ of each relay $i$ at time $t$. Table \ref{tab_SS} summarizes this state space representation. Note that the state  includes the past $m$ measurements.
\begin{table}
\begin{center}
	\caption{Relay State Space}
		\begin{tabular}{|c||c|}
			\hline
			\label{tab_SS}
			State Variable & Description\\
			\hline
			Voltage & \makecell{Voltage measurements of past $m$ timesteps}\\
			\hline
			 Current & \makecell{Current measurements of past $m$ timesteps}\\
			\hline
			Status & \makecell{Current breaker state: open or close}\\
			\hline
			Counter & Current value of the time counter\\
			\hline
		\end{tabular}
	\end{center}
\end{table}

To define the action space, we first specify the possible actions each relay can take. When a relay detects a fault it will  decide to trip. However, to facilitate the coordination between the network of relays, rather than tripping instantaneously, it will trigger a counter with a time countdown, indicating the relay will trip after certain time steps. If the fault is cleared by another relay during the countdown, the relay will  reset the counter to prevent mis-operation. Table \ref{tab_AS} summarizes the action space of each relay.  So, the action of relay $i$ at time  $t$, $a_{i,t}$, is one of these 11 possible values.  
\begin{table}
	\begin{center}
		\caption{Relay Action Space}
		\begin{tabular}{|c||c|}
			\hline
			\label{tab_AS}
			Action & Description\\
			\hline
			Countdown & Continue the counter\\
			\hline
			Set & Set counter to value 1 - 9\\
			\hline
			Reset & Stop activated counter\\
			\hline
		\end{tabular}
	\end{center}
\end{table}

The reward given to each relay is determined by its current action and fault status. A positive reward occurs  if,  i) it remains closed during normal conditions,  ii) it correctly operates after a fault in its assigned region or in its first downstream region when the corresponding primary relay fails. A negative reward is caused by, i) tripping when there is no fault; 2) tripping after a fault outside its assigned region.  The magnitude of the rewards are designed in such a way to facilitate the learning, implicitly signifying relative importance of false positives and false negatives. The reward function for each relay is shown in Table  \ref{tab_RW}. 
\begin{table}
	\begin{center}
		\caption{Rewards}
		\begin{tabular}{|c||c||c|}
			\hline
			\label{tab_RW}
			Condition & Trip & Hold\\
			\hline
			Normal & -150 & +3 \\
			\hline
			\makecell{After fault in \\main region} & +120 & \makecell{-3 for each post-fault step}\\
			\hline
			\makecell{After fault \\as backup} & +100 & \makecell{-2 for each post-fault step}\\
			\hline
			\makecell{After fault \\outside assigned region} & -150 & +5\\
			\hline
		\end{tabular}
	\end{center}
\end{table}

Consider a network with $n$ relays. Define the global state of the network at time $t$ as $\bar{s}_{t} = (s_{1,t}, s_{2, t}, \ldots, s_{n,t})$ and the global action at time $t$ as $\bar{a}_{t} = (a_{1,t}, a_{2, t}, \ldots, a_{n,t})$. Let $R_{i,t}$ be the reward obtained by relay $i$ at time $t$. It is clear from  the description of the system  that $R_{i,t}$ depends on the global state $\bar{s}_{t}$ and global action $\bar{a}_{t}$ rather than the local state ${s}_{i,t}$ and local action ${a}_{i,t}$ of relay $i$. Define the global reward $\bar{R}_{t}$ as $\bar{R}_{t} = \sum^{n}_{i=1} R_{i,t}$.  Note that the (global) state evolution of the network can longer be described by looking at the local transition probabilities because the control actions of the relays affect each others' states. The global dynamics is represented by the transition probability  $\bar{P} = \bar{P}(\bar{s}_{t+1} | \bar{s}_{t}, \bar{a}_{t})$. 

We formulate the  optimal relay protection problem in a network as multi-agent RL problem. The goal is to achieve a global objective, maximizing the cumulative reward obtained by all relays, using only local control laws $\pi_{i}$ which maps the local observations $s_{i,t}$ to local control action $a_{i,t}$.  Formally,
\begin{align}
\label{eq:marl1}
\max_{(\pi_{i})^{n}_{i=1}} ~ \mathbb{E}[\sum^{\infty}_{t=0} \gamma^{t} \bar{R}_{t} ],~ a_{i,t} = \pi_{i}(s_{i,t}).
\end{align}
Since the model is unknown and there is no communication between relays,  each relay has to learn its own local control policy $\pi_{i}$ using an RL algorithm to solve \eqref{eq:marl1}.

Classical RL algorithms and their deep RL versions typically address only the single agent learning problem. A multi-agent learning environment violates one of the fundamental assumption needed for the convergence of RL algorithms, namely, the stationarity of the operating environment. In a single agent system, for any fixed policy of the learning agent, the probability distribution of the observed states can be described using a stationary Markov chain.  RL algorithms are designed to learn only in  such a stationary Markovian environment. Multiple agents taking actions simultaneously violate this assumption. Even if the policy of a given agent is fixed,  state observations for that agent are  no longer according to a stationary Markov chain,  as they are controlled by  the actions of other agents.  Moreover, in our setting, each relay observes only its local measurements which further complicates the problem. A formulation that involves these two constraints is known as Decentralized Partially Observable Markov Decision Process (Dec-POMDP) \cite{MARLReport}. There are existing literatures \cite{MARLKraemer} addressing this kind of problems, but the performance of most algorithms are unstable and the convergence is rarely guaranteed. 

We propose an approach to overcome this difficulty of multi-agent RL problem by exploiting the radial structure of power distribution systems. Using this structural  insight, we develop a nested RL algorithm to extend the single agent RL algorithm to the multi-agent setting we address.   

We use the following training procedure. We start from the very end of the radial network in Fig. \ref{fig:radial1}. The relay protecting the last segment is relay 5, which has no downstream neighbors and can be trained using the single-agent training algorithm described in the previous section. Once the training of relay 5 is complete, it will react to the system dynamics using its learned policy. Since relay 5 only needs to clear local faults (i.e. faults between bus 862 - 838) and ignores disturbances at any other location, its policy will not change according to the change in the policy of other relays. This enables us to train relay 4 with relay 5 operating with a fixed policy (which it learned via the single agent RL algorithm). Since the policy of relay 5 remains fixed when training relay 4, the environment from the perspective of relay 4 remains more or less stationary (except for the possible disturbances due to difference in the local measurements). Similarly, after the training of relay 4 and 1 is complete, relay 2 can be trained with the policy of relay 1, 4 and 5 fixed. This process can be repeated for all the relays upstream to the feeder. 

This nested training approach which exploits the nested structure of the underlying physical system allows us to overcome the non stationarity issues presented in generic multi-agent RL settings. Our nested RL algorithm is formally presented in Algorithm \ref{algo:marl}.

\begin{algorithm}[ht]
	\caption{Nested RL for Radial Relay Network}
	\label{algo:marl}
	\begin{algorithmic}
		\STATE Sort $\{i|1\leq{i}\leq{n}\}$ into a vector $N$ by the ordering of training 
		\STATE Initialize replay buffer of each relay $N_i, 1 \leq i \leq n$ 
		\STATE Initialize Q-value function of each relay $i$ with random network weights $w_{i}$		
		\FOR {relay $i = 1$ to $n$} 
		\FOR {episode $k = 1$ to $M$}  
		\STATE Initialize PSS/E simulation environment  with randomly generated system  parameters
		\FOR {time step $t = 1$ to $T$} 
		\STATE Observe the state  $s_{i, t}$ of each relay $N_i$		
		\FOR {relay  $j = 1$ to $i$}
		\STATE With probability $\epsilon$ select a random action $a_{j,t}$, \\
		otherwise select a greedy action \\ $a_{j,t} = \arg \max_{a} Q_{w_{j}}(s_{j,t}, a)$
		\STATE Observe the reward $R_{j,t}$ and  next state $s_{j,t+1}$
		\STATE Store $(s_{j,t}, a_{j,t}, R_{j,t}, s_{j,t+1})$ in the replay buffer of relay $N_j$
		\STATE Sample a minibatch from replay buffer and update $w_{j}$
		\ENDFOR
		\FOR {relay  $j = i+1$ to $n$}
		\STATE Select the null action, $a_{j,t} = 0$
		\ENDFOR
		\ENDFOR
		\ENDFOR
		\ENDFOR
	\end{algorithmic} 
\end{algorithm}

%
%

\section{Simulation Results}  

In this section we evaluate the performance of our   RL algorithm  for protective relays. We compare its performance with the conventional threshold based relay protection strategy. We compare the performance in the following metrics: 

\noindent \emph{Failure rate:} We evaluate the  percentage of operation failures of relays in four different scenarios: when there is a: (i) fault in the local region, (ii) fault in the immediate downstream region, (iii) fault in a remote region, (iv) no fault in the network.   

\noindent \emph{Robustness to changes in the operating conditions:} Relays are trained for a given operating condition, like a specific load profile. We evaluate   protective relay strategies when there are changes in such operating conditions.  

\noindent \emph{Response time:} Relay protection control is supposed to work immediately after a fault occurs. We evaluate the time taken between the occurrence of a  fault and activation of the protection control.

\subsection{Simulation Environment Implementation}  
We choose the network structure shown in Fig. \ref{fig:IEEE34} for simulations. In particular, we focus on the relays in the section of the network shown in Fig. \ref{fig:radial1}. We first describe the simulation environment created for training and testing different relay protection strategies.  We implemented the environment using   {Power System Simulator for Engineering} (PSS/E) by Siemens, a commercial power system simulation software. The simulation process is controlled by Python using the official PSSPY interface library and the dynamic simulation module. The environment is wrapped according to the OpenAI Gym \cite{Gym} format that provides APIs for agents to start, step through and end an episode.      

An episode is defined as a short simulation segment that contains a fault. In the beginning of each episode, a random initial  operating condition (e.g. generator output, load size) is  generated  to mimic the load variation in distribution systems. During an episode, a fault is added to the system at a random time-step. The fault is set to have random fault impedance within a range determined according to the proposed model in \cite{Andrade2010} and occurs at a random location. In order to mimic the real world scenario that necessitates the backup relays,  a small probability to ignore a tripping action is added  for each relay. This corresponds to the case when the breaker fails as a relay tries to trip the line. In this case, the first upstream relay needs to trip the line instead.

In practice, the load level of distribution systems varies a lot with the time of day. To simulate this, the system is assigned a random trend factor from 70\% to 130\% in the beginning of each episode. In addition to this, each load in the system has its own load multiplier between 80\% and 120\% of the system-wide trend factor. The capacity of each load is determined by multiplying the base case capacity and the local multiplier. A powerflow solution is then calculated using this random load profile as the initial condition for the dynamic simulation.   

\subsection{RL Algorithm Implementation and Training}
The RL algorithm is implemented using the open-source library Keras-RL \cite{kerasRL}.  Since the Python interface of PSS/E used for the environment simulations and the  Keras-RL used for training the RL agents  are not compatible, they must run on two separate processes simultaneously. In order to enable this, the environment  is wrapped  as a TCP/IP server that accepts connections through a pre-allocated port and keeps running in the background. RL agents communicate with the server using the same port via $localhost$. Also, to accommodate the TCP/IP requirement, a dedicated encoder/decoder pair is also written to pack the data as byte strings before transmitting through the port.    We implemented Algorithm \ref{algo:marl} using this setup. Hyperparameters are selected via random search. The final configuration and hyperparameters of the RL algorithm are specified in Table \ref{tab_DQN}. All agents use the same hyperparameters in training.

\begin{table}
	\begin{center}
		\caption{DQN Agent Hyperparameters}
		\begin{tabular}{|c||c|}
			\hline
			\label{tab_DQN}
			Hyperparameter & Value\\
			\hline
			\makecell{Hidden Layers} & 2\\
			\hline
			Layer Size & 128/64\\
			\hline
			Activation & ReLU/ReLU/Linear \\
			\hline
			Replay Buffer & 10000\\
			\hline
			Replay Batch & 32\\
			\hline
			\makecell{Target Soft \\ Update Rate} & 0.005\\
			\hline
			Optimizer & Adam, Learing Rate 0.0005\\
			\hline
			Double DQN & On\\
			\hline
			Loss & MSE\\
			\hline
			Discount & 0.95\\
			\hline
		\end{tabular}
	\end{center}
	
\end{table}

The learning curves are shown in Fig. \ref{fig:lc1}. In particular, Fig. \ref{Fig3a} shows the convergence of episodic reward (cumulative reward obtained in an episode) for relay 5. The thick line indicates the mean value of episodic reward obtained during a trial consisting of 20 independent runs of training. The shaded envelope is bounded by the mean reward $\pm$ standard deviation recorded at the same progress during the trial. Note that the episode reward converges in less than 250 episodes. One episode takes roughly 3 seconds. So, the training converges in less than 750 seconds. 

Similarly, Fig. \ref{Fig3b} shows the convergence of the false operations for relay 5. In the beginning of training, the false operation rate is really high but it soon converges to a value approximately zero. Fig. \ref{Fig3c} shows the learning curve corresponding to the episodic reward of relay 4. The behavior is similar to that of relay 5 though it takes more number of episodes to converge. This is expected because relay 4 has to act both as primary relay and as backup for relay 5, while relay 5 only has to act as the primary relay (c.f. Fig. \ref{fig:radial1}). So, the control policy of relay 4 is more complex than the policy of relay 5, and hence it takes a longer training time to converge. Also, the standard deviation of episodic reward of relay 4 is slightly higher than relay 5. This is because the reward relay 4 can get in backup relay mode can be smaller than in primary relay mode even for perfect operation. 

We omit the learning curves for other relays as they are very  similar to that of relay 5 and relay 4.

\subsection{Conventional Relay Protection Strategy}

\begin{figure*}[t]
	\captionsetup[subfigure]{justification=centering}
	\begin{subfigure}[t]{1.9in}
		\centering
		\includegraphics[scale=0.33]{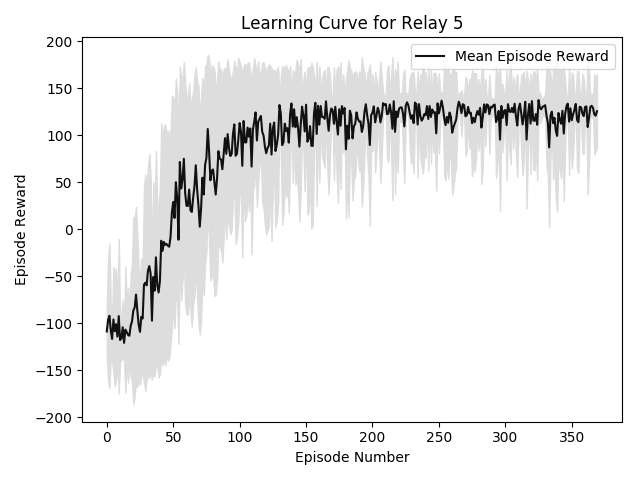}
		\caption{Relay 5: Episodic reward}\label{Fig3a}
	\end{subfigure}
	\begin{subfigure}[t]{1.5in}
		\centering
		\includegraphics[scale=0.33]{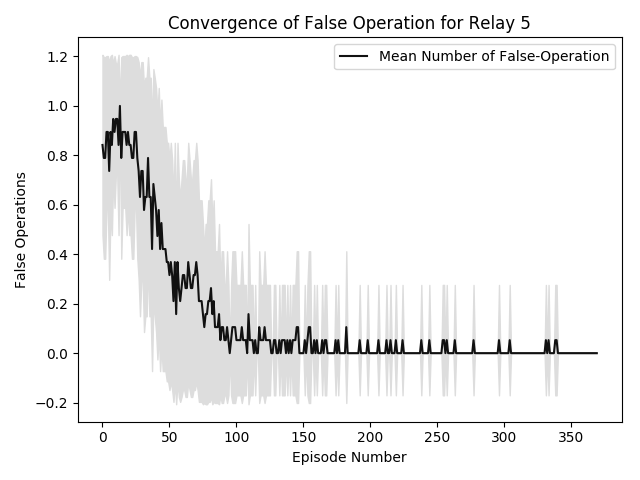}
		\caption{Relay 5: Failure Rate}\label{Fig3b}
	\end{subfigure}
	\begin{subfigure}[t]{2.5in}
		\centering
		\includegraphics[scale=0.33]{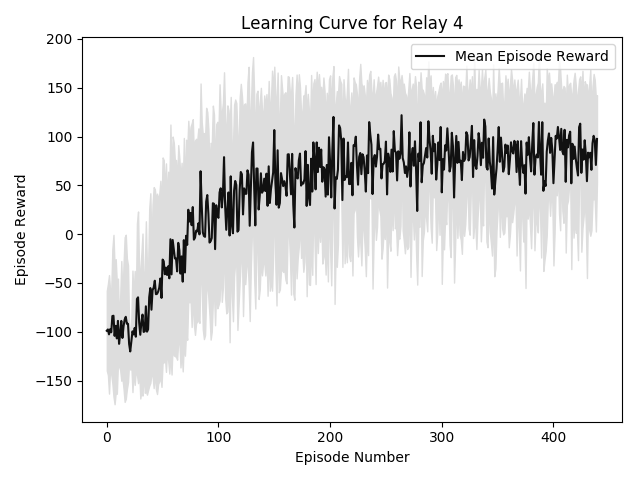}
		\caption{Relay 4: Episodic reward}\label{Fig3c}
	\end{subfigure}
	\begin{subfigure}[t]{2.3in}
		\centering
		\includegraphics[scale=0.35]{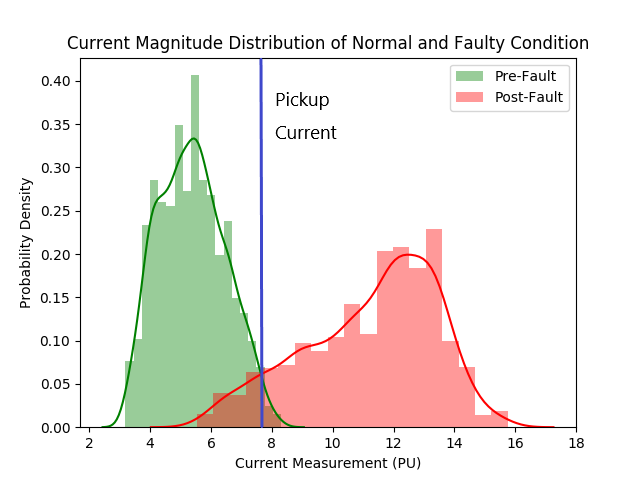}
		\caption{Normal/faulty current distribution}\label{Fig:pdf}
	\end{subfigure}
	\begin{subfigure}[t]{2.3in}
		\centering
		\includegraphics[scale=0.36]{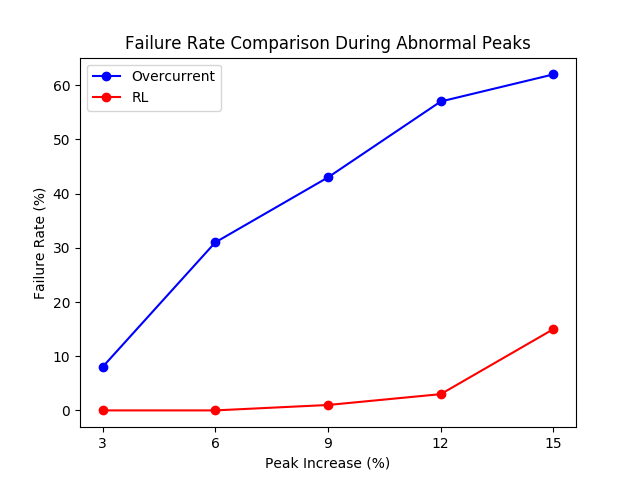}
		\caption{Failure Rate During Abnormal Peaks}\label{Fig3e}
	\end{subfigure}
	\begin{subfigure}[t]{2.0in}
		\centering
		\includegraphics[scale=0.36]{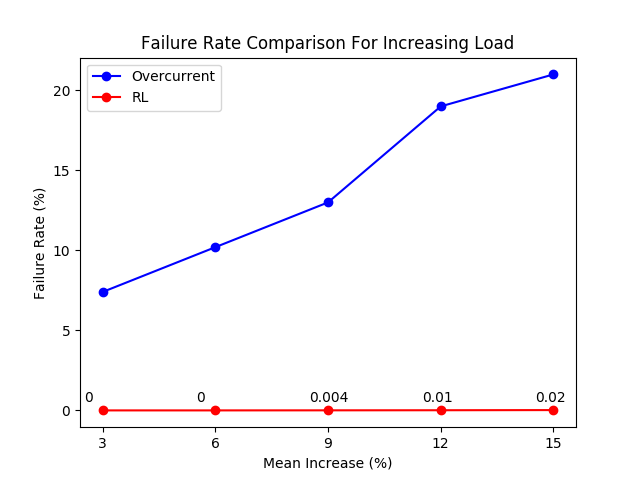}
		\caption{Failure Rate During Increased Mean Load}\label{Fig3f}
	\end{subfigure}
	\caption{Convergence Plots of Agents and Comparison of Robustness}
	\label{fig:lc1}
\end{figure*}
	
Conventional relay protection strategies are based on a fixed threshold rule. The relay trips if and only if the measured current is greater than a fixed threshold. The optimal threshold is typically computed using a variety of heuristic methods \cite{PRBook}. Since these methods depend on the network parameters like  topology, feeder capacity and load size, they may not give the optimal threshold that maximizes the success rate in our setting. So, for a fair comparison with a more powerful RL based algorithm, we compute the  threshold that guarantees the optimal performance using a simple statistical approach.

We compute the empirical probability distribution (pdf) of the current measurements before and after the fault from 500 episodes. For example, the pdf of the pre-fault and post-fault current at bus 862 is plotted in Fig. \ref{Fig:pdf}.  It is clear from the figure that the distributions of the pre-fault and post-fault currents overlap with each other. This is expected, especially in the power distribution systems, where the load profile varies greatly with the time of day. In fact, when the system is lightly loaded, the fault current with a relatively large fault impedance can be smaller than the normal operating current when the system is heavily loaded. We put a higher weight on faulty scenarios to overcome the imbalanced prior probabilities. The optimal threshold that will maximize the success rate can then be approximated as the crossing point of these two pdfs \cite{Bayes}. This point is marked as the `pickup current' in the figure, and is used as the threshold for the conventional relay protection strategy.

\subsection{Performance Evaluation}

In this section we compare the performance of the RL based relay protection strategy and threshold based conventional relay protection strategy. As mentioned above, we focus on three metrics of performance, namely, failure rate, robustness, and response time.  

\noindent {\bf Failure rate}: A \textit{false operation} of a relay is the one operation where that relay fails to operate as it supposed to do. There are two kinds of false operations, false-negative and false-positive. A false-positive happens if: (i) relay trips  when there is no fault, (ii) relay trips even if the location of fault is outside of the relay's assigned region,  (iii) backup relay trips before the primary relay.  A false-negative happens if: (i) relay fails to trip even if the location of the  fault is inside its assigned region, (ii) backup relay fails to trip even after its immediate downstream relay has failed. 

For the RL based algorithm, we use the parameters obtained after the training.  For the conventional relay strategy, we use the optimal threshold computed as described before. We test the performance in four scenarios. Each scenario is tested with 5000  episodes.  Failure rate is calculated as the ratio of the number of episodes with  failed operations to the total number of episodes. Failure rate comparison is given in Table \ref{tab_RES}. Note that our RL based algorithm remarkably outperforms the conventional relay strategy. For example, in the local fault scenario, the conventional strategy has a failure rate of $7.7\%$ where as our RL algorithm has a mere $0.26\%$. Also note that in two scenarios, backup and no fault, even after 5000 random episodes, RL based strategy didn't cause any operation failure. So, we put the failure rate as zero.

\begin{table}[h!]
	\caption{False Operation Rate Comparison}
	\begin{center}
		\begin{tabular}{|c||c||c||c|}
			\hline
			\label{tab_RES}
			\multirow{2}{*}{Scenario} & \multirow{2}{*}{Expected Operation} & \multicolumn{2}{|c|}{Failure Rate} \\
			& & Conventional & RL-based \\
			\hline
			Local Fault & Trip & 7.7\% & 0.26\% \\
			\hline
			Backup & Trip & 9.6\% & 0\% \\
			\hline
			Remote Fault & Hold & 3.8\% & 0.08\% \\
			\hline
			No Fault & Hold & 1.8\% & 0\% \\
			\hline
		\end{tabular}
	\end{center}
\end{table}

\noindent {\bf Robustness}:
Load profiles in a distribution system is affected by many events like weather, social activities, renewable generation, and electric vehicles charging schedules.  These events can generally cause the peak load to fluctuate and possibly exceed the expected range in the planning stage. Moreover, the electricity consumption is expected to slowly increase each year, reflecting the continuing economy and population growth \cite{EIAReport}. This can cause a shift in the mean (and variance) of the load profile. Relay protection control should be robust to such changes as continually reprogramming relays after deployment  is costly.

We first evaluate the robustness  in the case of peak load variations. For the clarity of illustration, we focus on relay 5.  We vary the peak load upto 15\% more than the maximum load  used during training. Since we are considering the robustness w.r.t. to the peak load variations, the load capacity used in this test is sampled only from peak load under consideration.  For example, the data collected for 3\% increase are sampled by setting the system load size between 100\% and 103\% of the peak load at training. We then test the performance of both relay strategies using the same parameter from the original training, i.e., we don't update the  policy to accommodate the change in this load profile. 

The performance is shown in Fig. \ref{Fig3e}. It can be seen that  conventional relay strategy completely fails against such a change in the operating environment as it fails in more than 40\% of such scenarios after a 9\% increase in the peak load. On the other hand, RL algorithm is remarkably robust at this point with failures in less than 2\% scenarios. RL algorithm performance starts to decay noticeably only after 15\% increase in the peak load. 

We also evaluate the robustness against increase in the mean load. Procedure is same as before and the performance  is shown in  Fig. \ref{Fig3f}.  RL algorithm is remarkably robust  even after a 15\% increase in the mean load. Conventional relay strategy fails completely in this scenario also.


\noindent {\bf Response time}: RL algorithm  also shows a very fast tripping speed during the testing. We observed a tripping time of $0.005$ second for the primary relay and $0.009$ second for the backup relay. Conventional overcurrent relays uses the time-inverse curves as the ones defined in IEEE C37.112-2018 \cite{IEEERelayStd} to determine the time delay for all situations, which gives unnecessary delay even for operations as primary relays. Depending on the curve selected, the minimum delay is at least at the order of 0.1 second.

\section{Concluding Remarks}

This paper proposes a multi-agent reinforcement learning based approach for redesigning the control architecture of protective relay in power distribution systems. We propose a novel nested reinforcement learning algorithm that exploits the underlying physical structure of the network in order to overcome the  difficulties associated with standard multi-agent problems. Unlike the generic multi-agent RL algorithms which often fail to converge, our nested RL algorithm converges fast in simulations. The converged policy far outperforms the conventional threshold based relay protection strategy in terms of failure rates, robustness to change in the operation conditions, and speed in responses. 

In the future work, we plan to develop analytical guarantees for the convergence of our nested RL algorithm. Future work will also investigate the scalability of the proposed work in much larger system and  the impact of network topology on the performance.

\end{document}